\documentclass{v16nufact}

\def\be{\begin{equation}}
\def\ee{\end{equation}}
\def\bea{\begin{eqnarray}}
\def\eea{\end{eqnarray}}

\newcommand{\Photo}{}
\def\gtap{\ \raise.3ex\hbox{$>$\kern-.75em\lower1ex\hbox{$\sim$}}\ }
\def\ltap{\ \raise.3ex\hbox{$<$\kern-.75em\lower1ex\hbox{$\sim$}}\ }
\newcommand{\ket}[1]{|{#1}\rangle}
\newcommand{\bra}[1]{\langle{#1}|}
\usepackage{wrapfig,epsfig}

\begin{document}
\vspace*{4cm}
\title{Recent Developments in Neutrino-Nucleus Scattering (Theory)}

\author{Satoshi X. Nakamura}

\address{
Department of Physics, Osaka University, Toyonaka, Osaka 560-0043, Japan}

\maketitle

\abstracts{
Selected theoretical developments in neutrino-nucleus scattering in
2015-2016 are reviewed.
}

\section{Introduction}

Next-generation neutrino oscillation experiments are going to 
address the leptonic CP violation and the neutrino mass
hierarchy,
and for this purpose, 
more accurate understanding of neutrino-nucleus reactions 
is needed.
Because the neutrino oscillation experiments 
utilize neutrino beams over a wide energy range,
the neutrino-nucleus reactions 
of different characteristics need to be understood.
From low to high energies,
the dominant reaction mechanism varies
from the quasi-elastic knockout of a
nucleon (QE), quasi-free excitation of the $\Delta(1232)$ or higher 
resonances followed by a decay into a meson-baryon final state (RES), 
and deep inelastic scattering (DIS).
A unified description of the neutrino-nucleus reactions over the wide
energy range needs to be developed;
recent efforts towards this direction are reported in Ref.~\cite{unified}.
In this review presentation,
I will cover selected theoretical developments in 2015-2016 in
neutrino-nucleus scattering in the QE and RES regions.

\section{QE}

The neutrino-nucleus scattering in the QE region is highly relevant to
the T2K experiment~\cite{t2k} that utilizes relatively low-energy neutrino beam
peaking at $\sim$ 0.6~GeV.
Recent theoretical interest in this subject has been to better
understand QE-like processes such as those involving meson-exchange
currents and final state interactions (FSI)~\cite{nieves_review}.
Here I will focus on a very recent update on the subject:
an ab initio calculation of inclusive electron scattering on
$^{12}$C in the QE region~\cite{lovato1,lovato2}. 
Ab initio calculations are presumably the best approach in
non-relativistic regime, apart from rather expensive computational
cost. 
Although this work is about the electron scattering, the same method
should work as well for neutrino scattering.
Also, the approach can be validated against
a large amount of precise electron scattering data.

In the ab initio approach to nuclear many-body problems,
one exactly (up to a certain numerical accuracy) solves the Schr\"odinger equation, 
$H\ket{\Psi_i} = E_i\ket{\Psi_i}$, 
where $E_i$ is an energy eigenvalue and $\ket{\Psi_i}$ is the
corresponding eigen-vector. 
The nuclear Hamiltonian is given by 
$H=\sum_i p^2_i/2 m_N + \sum_{i<j} v_{ij} + \sum_{i<j<k} v_{ijk} + \cdots$, 
where the first, second, and third terms are the kinetic term, $NN$
potential, and $3N$ potential, respectively. 
This nuclear many-body problem has been solved 
with the Green's function
Monte Carlo (GFMC) method, which is one of ab initio calculational methods,
and the result excellently reproduced 
data for ground and low-lying excited state energies
of light nuclei up to and including $A=12$ 
($A$: mass number)~\cite{spectra}.
This state-of-the-art technology has been applied to the inclusive
electron scattering~\cite{lovato1,lovato2}. 

The electron scattering is induced by electromagnetic current operators
acting on a nuclear wave function. 
The currents used in Refs.~\cite{lovato1,lovato2}
consist of one-body impulse current and two-body
meson-exchange currents.
The currents are constrained by the current conservation, and also by
data such as magnetic moments and electromagnetic form factors of light nuclei. 
With the transverse (longitudinal) electromagnetic current operators,
$J_T$ ($J_L$), the corresponding response function is defined by
\begin{eqnarray}
 R_\alpha(\omega,\vec q) = \sum_f \bra{\Psi_0}J^\dagger_\alpha(\omega,\vec q)\ket{\Psi_f}
\bra{\Psi_f}J_\alpha(\omega,\vec q)\ket{\Psi_0}\delta(\omega+E_0-E_f) \ , 
\quad \alpha=L,T 
\end{eqnarray}
where $\omega$ ($\vec q$) denotes the energy (momentum) transfer from
the electron to the nucleus;
$\Psi_0$ ($\Psi_f$) is the ground (an excited) state with the
corresponding energy $E_0$ ($E_f$).
The cross sections for the inclusive electron scattering on the nucleus
can be expressed with the response functions that
encode information of the nuclear dynamics.

\begin{wrapfigure}[32]{r}{0.6\textwidth}
   \begin{center}
\vspace{-3mm}
\includegraphics[width=0.58\textwidth]{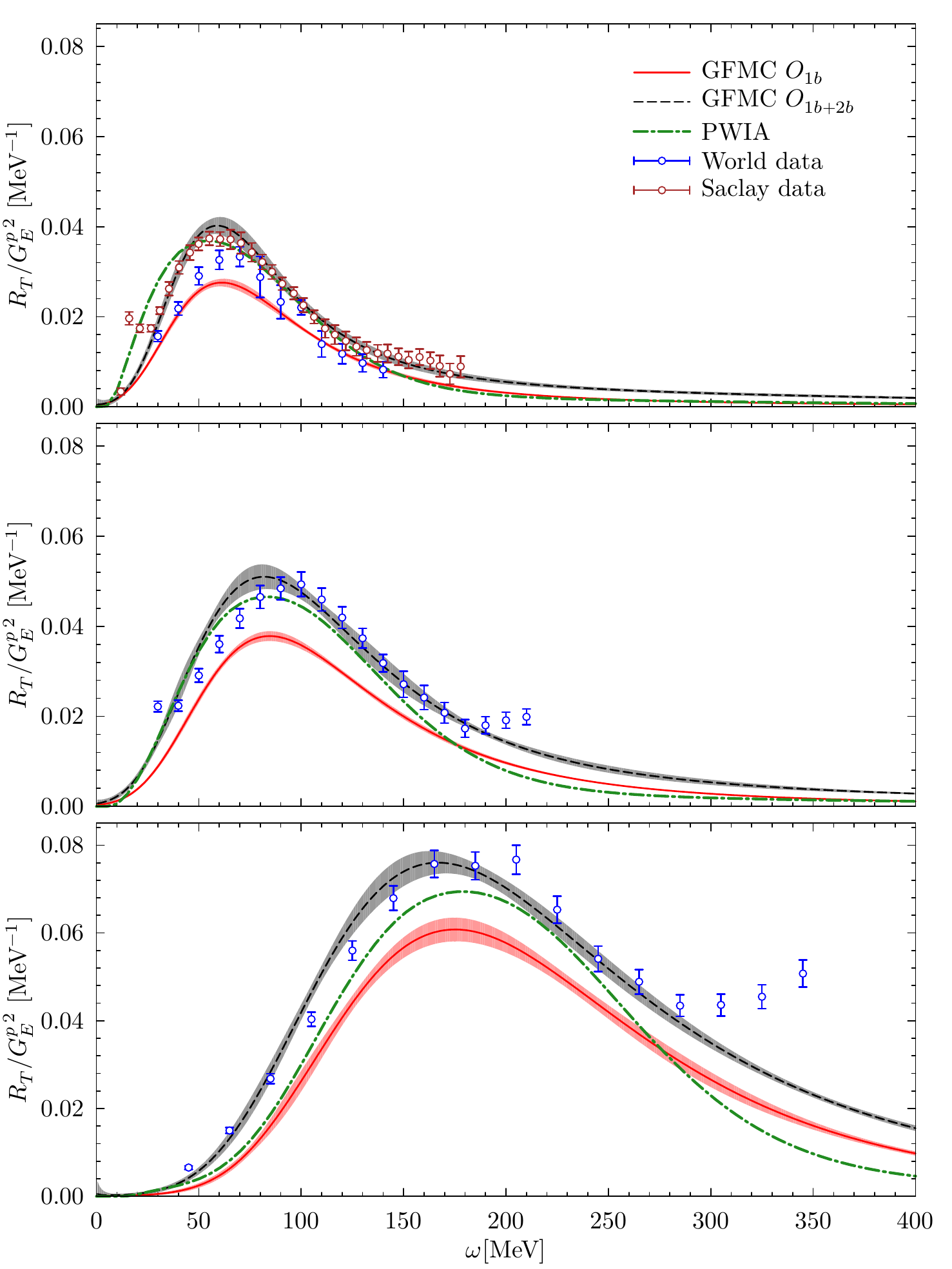}
   \begin{minipage}{0.52\textwidth}
\caption{\label{fig:c12} (Color online)
Transverse response functions for inclusive electron scattering on $^{12}$C.
The top, middle, and bottom figures are for $|\vec q|=$300, 380, and
    570~MeV, respectively.
Data are from the world data analysis$^7$ and from Saclay$^8$.
Figures taken from Ref.$^5$. Copyright (2016) APS.
}
\end{minipage}
   \end{center}
\end{wrapfigure}
The authors of Refs.~\cite{lovato1,lovato2} calculated the response functions for
$^{12}$C. Because a direct evaluation of the response functions is
formidably difficult, they took a strategy to first evaluate the
Laplace transform of $R_\alpha(\omega,\vec q)$ with the GFMC method, and
then invert it to obtain $R_\alpha(\omega,\vec q)$;
for the inversion, the maximum entropy method was employed. 
The obtained transverse response, $R_T(\omega,\vec q)$, 
divided by the square of $G^p_E(\omega,\vec q)$ (the electric
form factor of the proton) is shown in Fig.~\ref{fig:c12} where
experimental data are also shown for a comparison.
The calculation done with both one- and two-body currents (only
one-body current) is labelled by 'GFMC $O_{1b+2b}$' ('GFMC $O_{1b}$')
in the figure.
It is clearly seen in the figure that
the response functions of GFMC $O_{1b+2b}$ agree with the data
excellently, and are significantly enhanced from those of
GFMC $O_{1b}$ over the whole $\omega$ region.
Also shown in Fig.~\ref{fig:c12} is 'PWIA' which is the plane-wave
impulse approximation and is calculated with the single nucleon momentum
distribution of $^{12}$C from Ref.~\cite{12C-single-N}.
The response functions of
GFMC $O_{1b}$ and those of PWIA are significantly different, indicating
a large reduction due to FSI.
These results validate the predictive power of the ab initio approach in
this kinematical regime. 
Other nuclear many-body models that involve 
approximations and/or truncations of model space should be validated by
either data or the ab initio calculations.
%

\section{RES}

The neutrino-nucleus scattering in the RES region is relevant to
low-energy ($E_\nu \ltap$ 1~GeV) experiments such as T2K~\cite{t2k} where 
a pion produced by $\Delta(1232)$ can be stuck in the nucleus, giving a
CCQE-like event. 
It is also relevant to
relatively high-energy
($E_\nu = 2\sim 4$~GeV) experiments such as DUNE~\cite{dune} where
higher resonances are excited to produce one or two pions.
For theoretically describing the processes, 
we need nuclear models that describe initial nucleon correlations, FSI,
and medium modifications of hadron properties.
We also need a model that describes neutrino reactions on a single
nucleon. Elementary amplitudes from the model are a building block to
construct a neutrino-nucleus scattering model. 
In what follows, I will focus on two recent works for developing
an elementary process model in the RES region. 

\subsection{Dynamical coupled-channels model for neutrino-induced meson productions}

In the RES region, 
particularly between the $\Delta(1232)$ and DIS regions,
developing a model for neutrino-induced meson productions off a single
nucleon is still an issue.
Several theoretical models have been developed, and
particularly
the $\Delta(1232)$ region has been extensively studied because of
its importance.
However, there still remain conceptual and/or practical problems in the existing
models as follows:
First, 
reactions in the RES region are
multi-channel processes in nature.
However, no existing model takes account of
the multi-channel couplings required by the unitarity.
Second, neutrino-induced double pion productions over the entire RES
     region have not been studied in detail previously,
even though their production rates are expected to be 
comparable or even more important than those for the single-pion productions
around and beyond the second RES region.
Third,
interference between resonant and non-resonant amplitudes are not 
well under control for the axial current in most of the
previous models.
In Ref.~\cite{dcc-neutrino},
the authors developed a neutrino-nucleon reaction model in the
RES region by overcoming the problems mentioned above;
this is what I will review in this subsection.

For developing a neutrino-nucleon reaction model in the
RES region, the authors of Ref.~\cite{dcc-neutrino}
took 
the best available option: working with a
coupled-channels model.
In the last few years, the authors' group
have developed a dynamical coupled-channels (DCC)
model to analyze $\pi N, \gamma p\to \pi N, \eta N, K\Lambda, K\Sigma$
reaction data for a study of the baryon spectroscopy~\cite{knls13}.
In there, they have shown that 
the model is successful in giving
a reasonable fit to a large amount ($\sim$~23,000 data points) of the
data in the RES region.
The model also has been shown to give a reasonable prediction for
pion-induced double pion productions~\cite{kamano-pipin}.
Thus the DCC model seems a promising starting point for developing a
neutrino-reaction model in the RES region.
For extending the DCC model to the sector of neutrino reactions,
the authors of Ref.~\cite{dcc-neutrino} made 
the following developments.
Regarding the vector current, they already had fixed the amplitude for the
proton target at $Q^2$=0 in their previous analysis~\cite{knls13}.
The remaining task was to determine the $Q^2$-dependence of the vector
couplings, i.e., form factors.
This was achieved by analyzing data for the single pion
electroproduction and inclusive electron scattering.
A similar analysis was also done for the neutron target data.
By combining the vector current amplitudes for the proton and neutron
targets, 
the isospin separation of the vector current amplitudes was made;
this is a necessary step for applying the vector current amplitudes to
the neutrino reactions.
Regarding the axial current, 
its matrix elements for tree-level non-resonant processes
were
derived from the chiral Lagrangian; the same Lagrangian has been used to
derive the $\pi N\to MB$ ($MB$: a meson-baryon state) potentials in the
DCC model.
By construction, the PCAC relation is satisfied.
Because of rather scarce neutrino reaction data, it is difficult to
determine 
$N$-$N^*$ transition matrix elements induced by the axial current.
The conventional practice is to write down a $N$-$N^*$ transition
matrix element induced by the axial current
in a general form with three or four form factors.
Then the PCAC relation is invoked
to relate the presumably most important axial form factor
to the corresponding $\pi NN^*$ coupling.
The other form factors are ignored except for the pion pole term. 
In Ref.~\cite{dcc-neutrino},
the axial currents for bare $N^*$ of
the spin-parity $1/2^\pm$, $3/2^\pm$, $5/2^\pm$ and $7/2^\pm$ were considered,
and the above procedure was taken to
determine their axial form factors at $Q^2=0$.
As a result of this derivation, 
the interference pattern between the resonant and non-resonant
amplitudes are uniquely fixed within their DCC model;
this is a great advantage of the DCC approach.
For the $Q^2$-dependence of the axial-current matrix elements,
a simple ansatz was taken inevitably
due to the lack of experimental information.
This is a limitation shared by all the existing neutrino-reaction models
in the RES region.
The $Q^2$-dependence of all the axial-coupling was 
assumed to be the conventional dipole form with 
the axial mass, $M_A=1.02$~GeV.

With the vector and axial currents as described above, 
cross sections for the neutrino-induced meson productions in the RES
region were calculated.
\begin{figure}[t]
\includegraphics[height=0.24\textwidth]{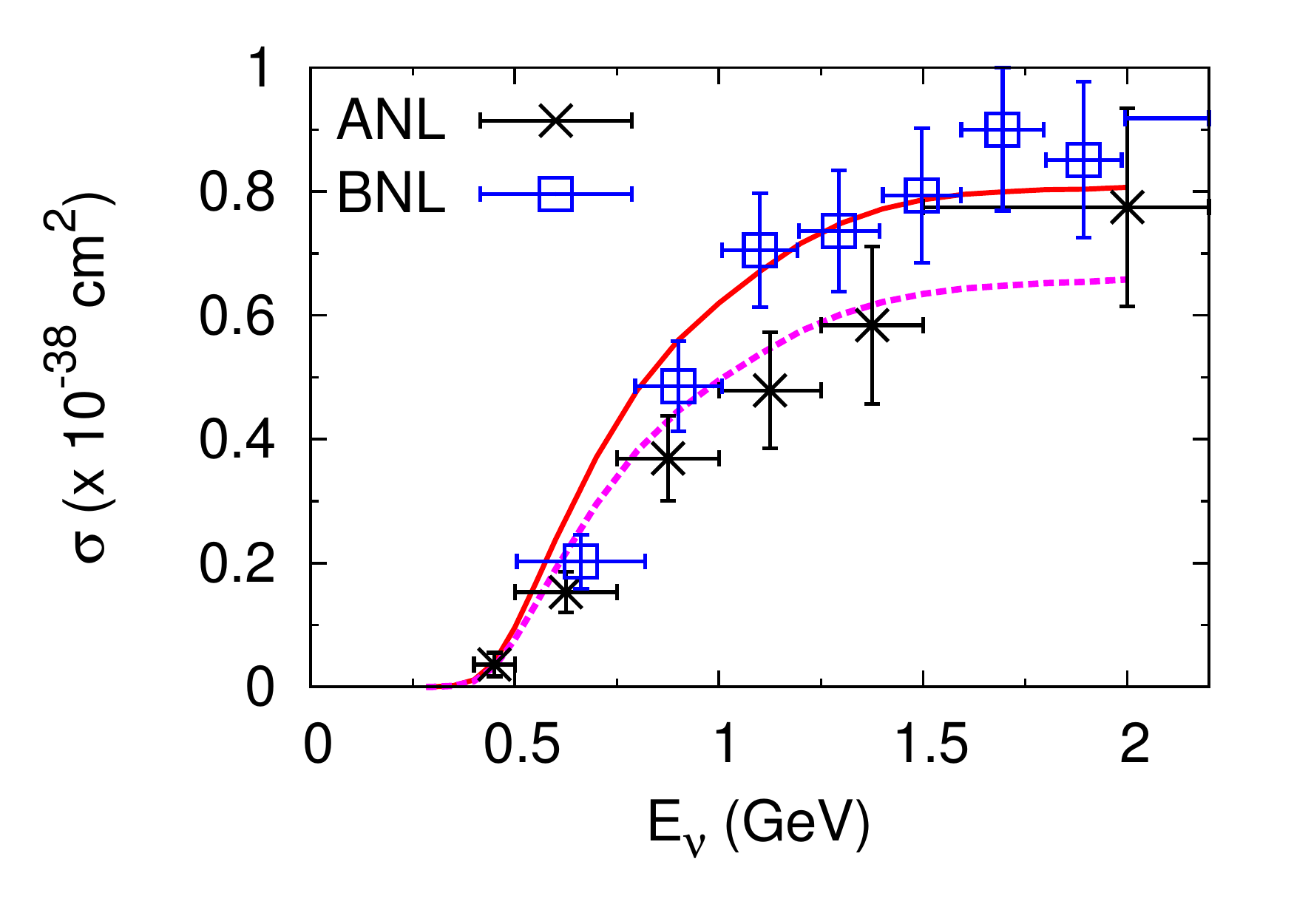}
\hspace{-5mm}
\includegraphics[height=0.24\textwidth]{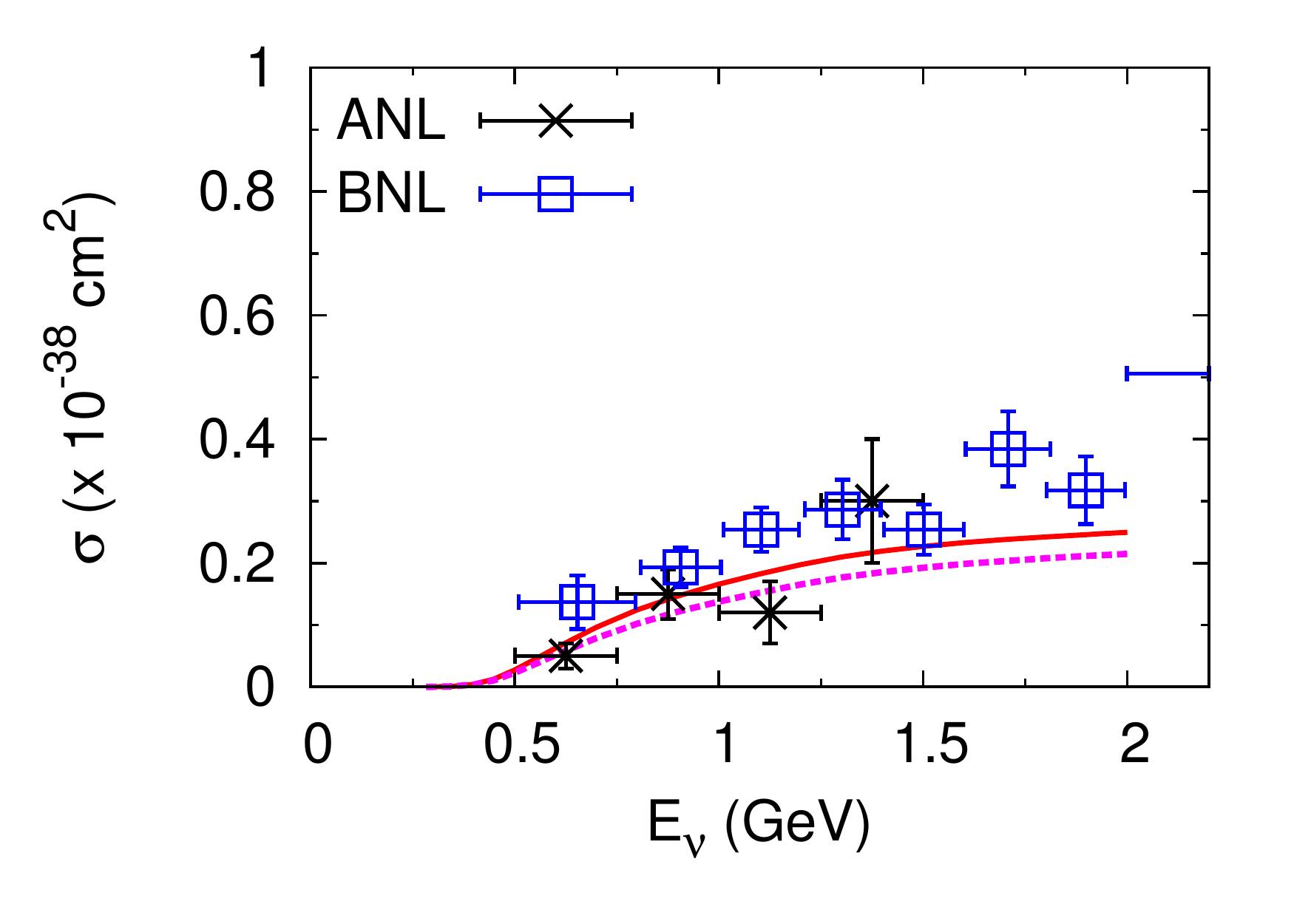}
\hspace{-5mm}
\includegraphics[height=0.24\textwidth]{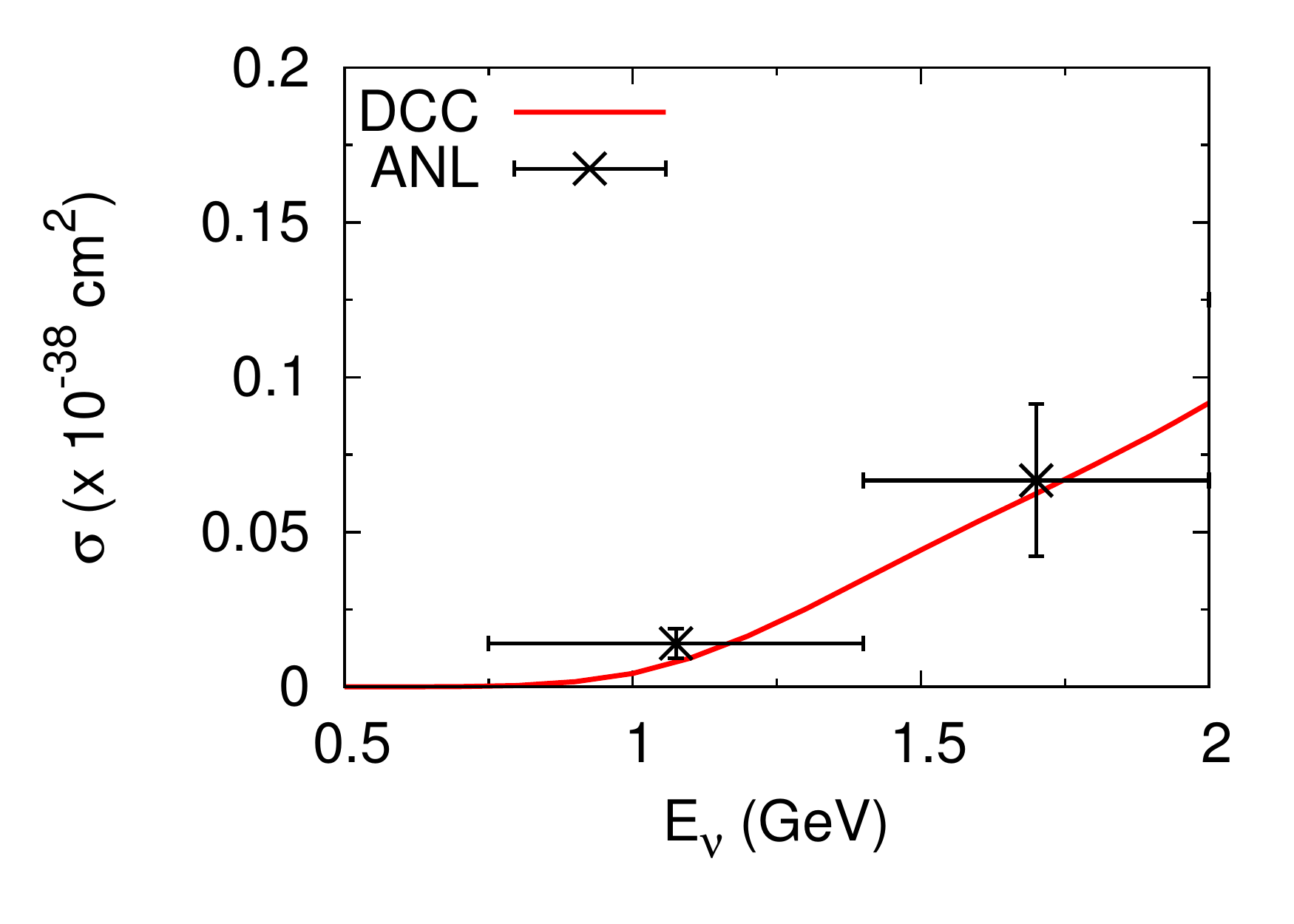}
\caption{(Color online)
Comparison of the DCC-based calculation (red solid curves)
with data for 
$\nu_\mu\, p\to \mu^- \pi^+ p$ (left),
$\nu_\mu n\to \mu^- \pi^0 p$ (middle)
and $\nu_\mu\, p\to \mu^- \pi^+\pi^0 p$ (right).
The DCC calculation with $0.8\times g_{AN\Delta(1232)}^{\rm PCAC}$
is also shown (magenta dashed curve).
The data are from ANL$^{14}$ and BNL$^{15}$ in the left and middle
panels, and ANL data are from Ref.$^{16}$ in the right panel.
Figures taken from Ref.$^{11}$. Copyright (2015) APS.
}
\label{fig:neutrino-tot-data}
\end{figure}
The calculated CC neutrino-induced single pion production cross
sections from the DCC model are compared with
available data from Refs.~\cite{anl,bnl} in
Fig.~\ref{fig:neutrino-tot-data} (left, middle).
The left panel shows the total cross sections for
$\nu_\mu\, p\to \mu^- \pi^+ p$ for which $\Delta(1232)$ dominates.
The two datasets from ANL~\cite{anl} and BNL~\cite{bnl}
for $\nu_\mu p\to \mu^- \pi^+ p$
shown in the left panel of
Fig.~\ref{fig:neutrino-tot-data} are not consistent
as has been well known,
and the DCC calculation is closer to the BNL data~\cite{bnl}.
For the neutron target (middle), the DCC calculation is fairly
consistent with both of the ANL and BNL data.
It seems that 
the bare axial $N$-$\Delta(1232)$ coupling constants
determined by the PCAC relation are too large to 
reproduce the ANL data.
Thus the bare axial
$N$-$\Delta(1232)$ coupling constants,
$g_{AN\Delta(1232)}^{\rm PCAC}$, is multiplied 
by 0.8 so that the DCC model better fits the ANL data.
The resulting cross sections are shown by the dashed curves in 
Fig.~\ref{fig:neutrino-tot-data} (left,middle).
We see that
$\sigma(\nu_\mu p\to \mu^- \pi^+ p)$ is reduced due to the
dominance of the $\Delta(1232)$ resonance in this channel, while 
$\sigma(\nu_\mu n\to \mu^- \pi N)$ is only slightly reduced.
The original data of these two experimental data have been
reanalyzed recently~\cite{reanalysis}, and it was
claimed that the discrepancy between the two datasets is resolved. 
The resulting cross sections are closer to the original ANL data.
It is noted that the data shown in Fig.~\ref{fig:neutrino-tot-data} were taken from
experiments using the deuterium target.
Thus one should analyze the data considering the nuclear effects such as
the initial two-nucleon correlation and the final state interactions.
In the next subsection, I review a recent work~\cite{wsl}
that took a first step toward such an analysis.

Next the DCC calculation for a double-pion production is 
compared with existing
data in Fig.~\ref{fig:neutrino-tot-data} (right).
Although there exist a few theoretical works on the neutrino-induced double-pion
production near threshold, 
this DCC calculation is the first one that took account of relevant resonance
contributions for this process.
The DCC-based prediction is 
in good agreement with the data for 
the $\nu_\mu\, p\to \mu^- \pi^+\pi^0 p$ cross sections.
For a fuller presentation of the DCC calculations, see Ref.~\cite{dcc-neutrino}.

\subsection{Effects of final state interactions on pion productions in
  neutrino-deuteron reactions}

The bubble chamber experiments at ANL~\cite{anl} and BNL~\cite{bnl}
measured cross sections for $\nu_\mu d\to \mu^-\pi NN$ ($d$: deuteron)
and, from the data, the elementary 
$\nu_\mu N\to \mu^-\pi N$ cross sections were extracted.
However, the FSI was not taken into account
in extracting the single nucleon cross sections.
This is disturbing 
because the data (cross sections for the elementary
processes)~\cite{anl,bnl} are essentially only available information to
determine the strength of the dominant 
nucleon-$\Delta(1232)$ transition induced by the axial-current.
Thus the uncertainty of the data~\cite{anl,bnl} is directly reflected in
uncertainty of theoretical calculations for neutrino-nucleus cross sections.
In this subsection, I review
a recent work~\cite{wsl}
that studied the FSI effect on 
the $\nu_\mu d\to \mu^-\pi NN$ reactions.

\begin{figure}[t]
\begin{center}
\includegraphics[height=0.28\textwidth]{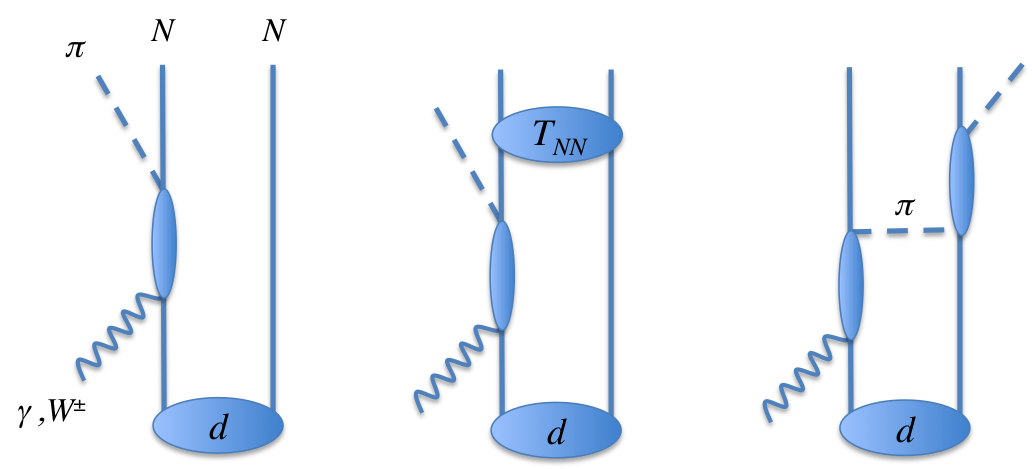}
\end{center}
\caption{
Mechanisms considered in Ref.$^{18}$
for describing $\gamma d\, (W^\pm d)\to \pi NN$ reactions.
(Left) Impulse term; (Middle) $NN$ rescattering term; 
(Right) $\pi N$ rescattering term.
}
\label{fig:nud}
\end{figure}

The authors of Ref.~\cite{wsl} used a dynamical model (called SL model)
developed in Ref.~\cite{sul} to generate elementary amplitudes that go
into a model for the neutrino-deuteron reactions.
The SL model has been shown to reproduce well 
the electromagnetic pion production data  in the $\Delta(1232)$ region, 
as well as the ANL and BNL data for
the neutrino-induced single pion production cross sections.
The incoherent electroweak pion productions on the deuteron 
were described in Ref.~\cite{wsl}
with the impulse term (Fig.~\ref{fig:nud} (left)),
the $NN$ rescattering term (Fig.~\ref{fig:nud} (middle)),
and the $\pi N$ rescattering term (Fig.~\ref{fig:nud} (right)).
In the diagrams of Fig.~\ref{fig:nud},
the $\gamma N\, (W^\pm N)\to \pi N$ and $\pi N\to \pi N$ amplitudes
are generated by the SL model, while the $NN$ scattering amplitudes and
the deuteron wave function are obtained with the Bonn potential~\cite{bonn}.

\begin{figure}[t]
\begin{minipage}[t]{75mm}
\begin{center}
\includegraphics[width=0.95\textwidth]{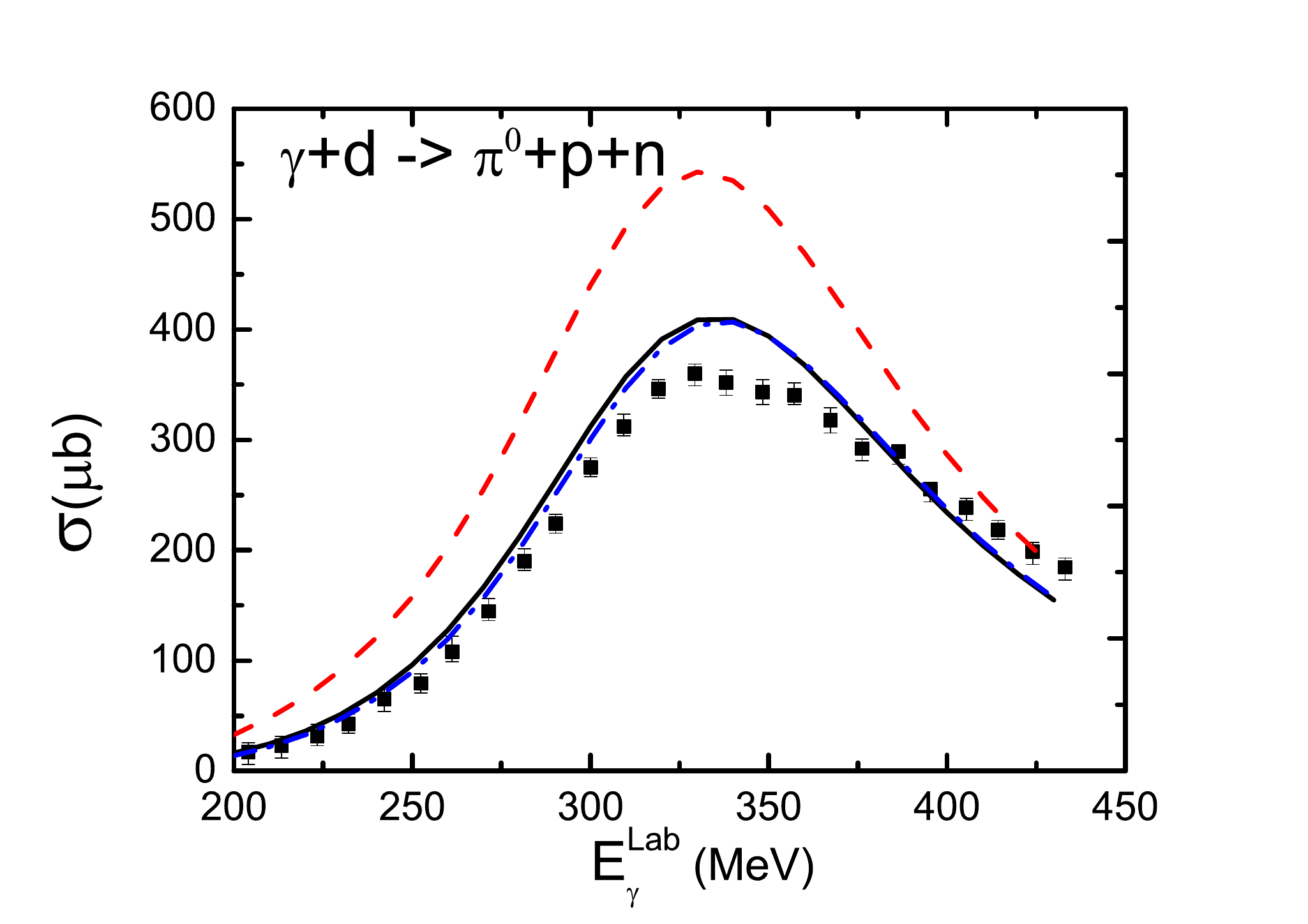}
\end{center}
\caption{(Color online)
FSI effect on total cross sections for 
$\gamma d\to \pi^0 pn$ reaction.
The red dashed curve is obtained with the impulse term only. 
By including the $NN$ ($NN$ and $\pi N$) rescattering terms, 
the blue dash-dotted (the black solid) curve is obtained. 
Figures taken from Ref.$^{18}$. Copyright (2015) APS.
}
\label{fig:EM}
 \end{minipage}
\hspace{5mm}
\begin{minipage}[t]{75mm}
\begin{center}
\includegraphics[width=0.95\textwidth]{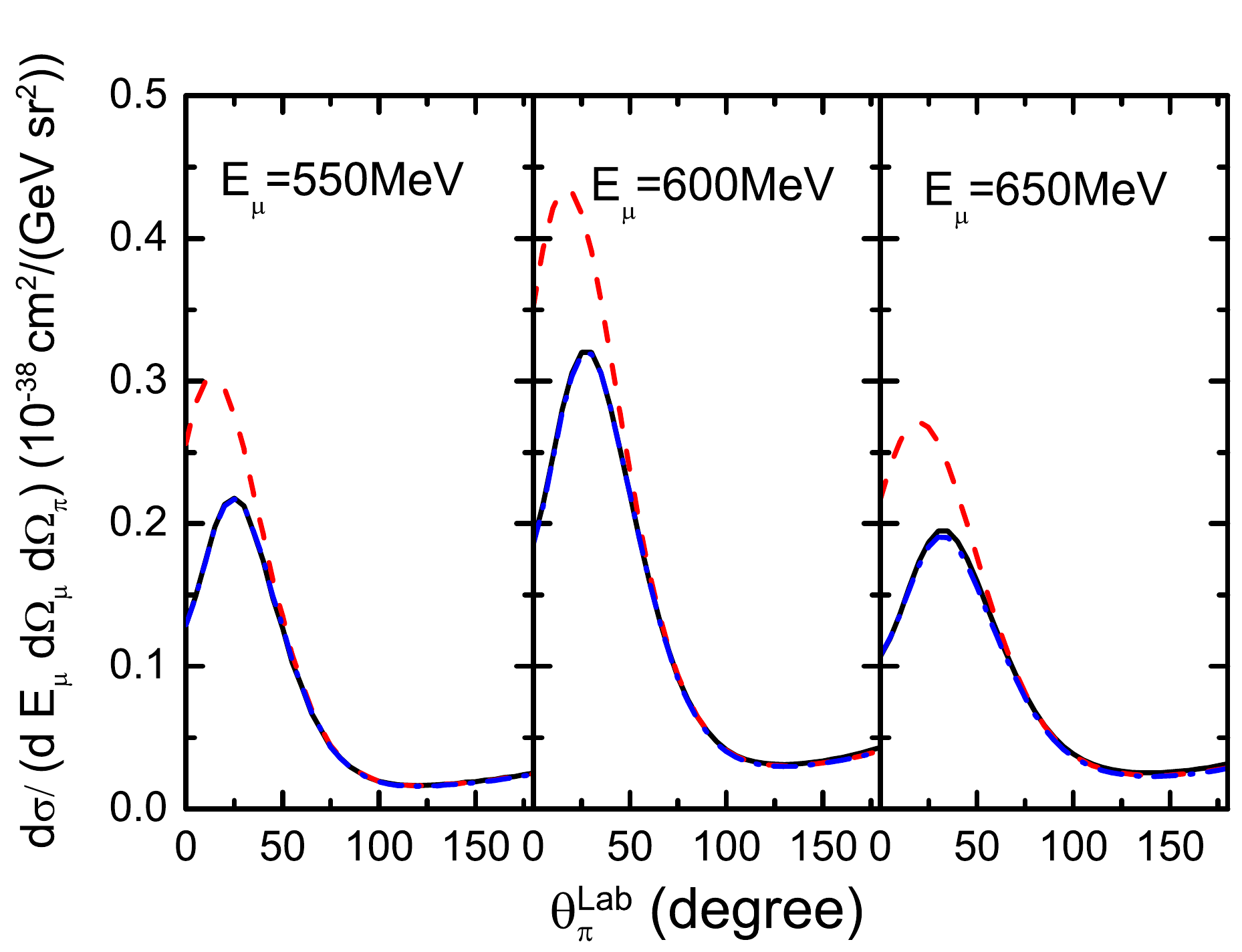}
\end{center}
\caption{(Color online)
FSI effect on differential cross sections for 
$\nu_\mu d\to \mu^-\pi^+ pn$ reaction at $E_\nu=$ 1~GeV,
 $\theta_\mu=25^\circ$, $\phi_\pi=0^\circ$.
The features of the curves are the same as those in Fig.~\ref{fig:EM}.
Figures taken from Ref.$^{18}$. Copyright (2015) APS.
}
\label{fig:fsi}
 \end{minipage}
\end{figure}

With the model described above, the authors of Ref.~\cite{wsl} first
studied pion photoproduction on the deuteron. 
Because there are precise and abundant data available for the
process, they can confront the model with the data to examine
the soundness of the model. 
Also, they can test the vector current with the photo-reaction data before
using it for neutrino processes.
Their calculation for $\gamma d\to \pi^0 pn$ total cross section
is shown in Fig.~\ref{fig:EM}.
The figure illustrates how the FSI modify the cross sections calculated
with the impulse term only. 
It is clearly seen that the $NN$ rescattering effect largely reduces the
cross sections, and brings the calculation much closer to the data.
This large reduction can be understood as a consequence of the
orthogonality between the deuteron and the $pn$ scattering wave
functions.
On the other hand, the $\pi N$ rescattering effect turned out to be
rather small.
The authors of Ref.~\cite{wsl} also found that the FSI effect 
is small for the $\gamma d\to \pi^- pp$ cross section where the
orthogonality does not work.

Having seen the capability of the model in the electromagnetic sector above, 
the model was then applied to the neutrino-induced pion productions. 
They chose a set of kinematical region where the cross section gets
large, i.e., the quasi-free $\Delta(1232)$-excitation kinematics.
Thus, for $E_\nu=$1~GeV, the muon kinematics is chosen to be 
$\theta_\mu=25^\circ$ and $E_\mu=$ 550, 600, 650~MeV.
The result for the differential cross section for 
$\nu_\mu d\to \mu^-\pi^+ pn$
is shown in Fig.~\ref{fig:fsi} as a function of the pion emission angle.
The $NN$ rescattering effect is again seen to be sizable.
This is because the orthogonality between the deuteron and scattering
$pn$ state is at work. 
The $\pi N$ rescattering effect is again rather small.
In this way, Ref.~\cite{wsl} showed that the FSI effect
is quite sizable in the neutrino-induced pion production on the deuteron,
although it had been
ignored in the previous analyses~\cite{anl,bnl} for extracting 
cross sections of the elementary processes.
The analysis of Ref.~\cite{wsl} was limited to a certain kinematics as
seen above, and thus they did not analyze the bubble chamber
data~\cite{anl,bnl} to extract cross sections of the elementary processes.
The data analysis requires integration all over the phase space,
which is rather demanding computationally.
Even so, considering that
accurate cross section data for the elementary
processes are highly demanded,
it would be important to
reanalyze the bubble chamber data with the sizable FSI effect taken into account.

\section*{Acknowledgments}

My participation to NUFACT 2016 was supported by JSPS KAKENHI Grant Number JP25105010.

\end{document}